\documentclass[english,twocolumn,superscriptaddress]{revtex4-2}
\usepackage[T1]{fontenc}
\usepackage[latin9]{inputenc}
\usepackage{amstext}
\usepackage{graphicx}
\usepackage{color, xcolor}

\makeatletter

\providecommand{\tabularnewline}{\\}

\makeatother

\usepackage{babel}
\begin{document}
\title{Neural network based time-resolved state tomography of superconducting qubits}
\author{Ziyang You}
\affiliation{Institute of Applied Physics and Materials Engineering, University
of Macau, Macau, China}
\author{Jiheng Duan}
\affiliation{Department of Physics and Astronomy, University of Rochester, Rochester,
New York 14627, USA}
\author{Wenhui Huang}
\affiliation{Shenzhen Institute for Quantum Science and Engineering, Southern University
of Science and Technology, Shenzhen, China}
\affiliation{International Quantum Academy, Shenzhen 518048, China}
\author{Libo Zhang}
\affiliation{Shenzhen Institute for Quantum Science and Engineering, Southern University
of Science and Technology, Shenzhen, China}
\affiliation{International Quantum Academy, Shenzhen 518048, China}
\author{Song Liu}
\affiliation{Shenzhen Institute for Quantum Science and Engineering, Southern University
of Science and Technology, Shenzhen, China}
\affiliation{International Quantum Academy, Shenzhen 518048, China}
\author{Youpeng Zhong}
\affiliation{Shenzhen Institute for Quantum Science and Engineering, Southern University
of Science and Technology, Shenzhen, China}
\affiliation{International Quantum Academy, Shenzhen 518048, China}
\author{Hou Ian}
\email{houian@um.edu.mo}

\affiliation{Institute of Applied Physics and Materials Engineering, University of Macau, Macau, China}
\begin{abstract}
Superconducting qubits have emerged as a premier platform for large-scale quantum computation, yet the fidelity of state readout is often hindered by random noise and crosstalk, especially in multi-qubit systems. While neural networks trained on labeled data have shown promise in reducing crosstalk effects during readout, their current capabilities are limited to binary discrimination of joint-qubit states due to architectural constraints. Here we introduce a time-resolved modulated neural network capable of full-state tomography for individual qubits, enabling detailed time-resolved measurements like Rabi oscillations. This scalable approach, with a dedicated module per qubit, mitigated readout error by an order of magnitude under low signal-to-noise ratios and substantially reduced variance in Rabi oscillation measurements. This advancement bolsters quantum state discrimination with neural networks, and propels the development of next-generation quantum processors with enhanced performance and scalability.
\end{abstract}
\maketitle

\section{Introduction}
Artificial intelligence (AI) has been increasingly utilized to address challenges across various domains of physics, including the field of quantum information science.
The integration of AI with quantum physics, initially explored in molecular electronics~\cite{Montavon}, has expanded to include AI-assisted measurements in various quantum systems, including quantum light~\cite{Tiunov,Ahmed},
trapped ions~\cite{Seif,Ding}, and superconducting qubits~\cite{Convy,Koolstra,Cao}.
Among these quantum systems, superconducting qubits have recently emerged as a leading platform for scalable quantum computation, offering remarkable coherence, controllability, and scalability~\cite{Arute,Clarke,Ladd}.
These properties have paved the way for high-fidelity single and two-qubit gates, aligning with the requirements for surface-code quantum error correction~\cite{google2023suppressing}.
However, the readout fidelity of superconducting qubits, particularly in simultaneous multi-qubit readouts, remains a significant hurdle~\cite{Borjans,Myerson,Robledo}. 
In the context of circuit quantum electrodynamics, qubit states are typically inferred by detecting the state-dependent frequency shift in a coupled resonator~\cite{Gambetta,Lupa=00015Fcu,Siddiqi}.
This measurement process is vulnerable to various noise sources, such as 1/f noise~\cite{Wang,Klimov}, flux noise~\cite{Martinis}, crosstalk~\cite{Tripathi,Mundada}, and the qubit's relaxation time~\cite{Pitsun,Walter}.

To address these challenges, machine learning techniques have been increasingly adopted~\cite{Heisoo,Boissonneault}, offering a non-invasive approach to enhance quantum measurement fidelity without necessitating physical modifications to the quantum system~\cite{Hornik,Cybenko}.
Neural networks, in particular, have demonstrated their efficacy in extracting meaningful features from raw data, bypassing the need for detailed error knowledge or reliance on predefined functions~\cite{Hornik,Cybenko}.
For superconducting qubits, neural networks have been employed to improve the fidelity of state discrimination between the ground and excited states~\cite{Duan,Lienhard,Magesan}.
They have also shown promise in analyzing joint states of multiple qubits. However, as the number of qubits increases, so does the complexity of the neural networks and the computational resources required. To date, neural networks have not been applied to perform complete qubit state tomography over extended periods, a task that requires a more sophisticated network architecture.

In this Letter, we introduce a time-resolved modulated neural network (TRMNN) specifically designed for the readout of single qubits in arbitrary superposition states. The TRMNN adopts a modular architecture, with each module featuring a fixed neural network, allowing for straightforward scalability by assigning one module per qubit. This design is optimized for continuous monitoring of the readout resonator, enabling comprehensive state tomography over user-defined durations, and facilitating the observation of Rabi oscillations at specific time intervals.

\begin{figure}
\includegraphics[width=8.6cm]{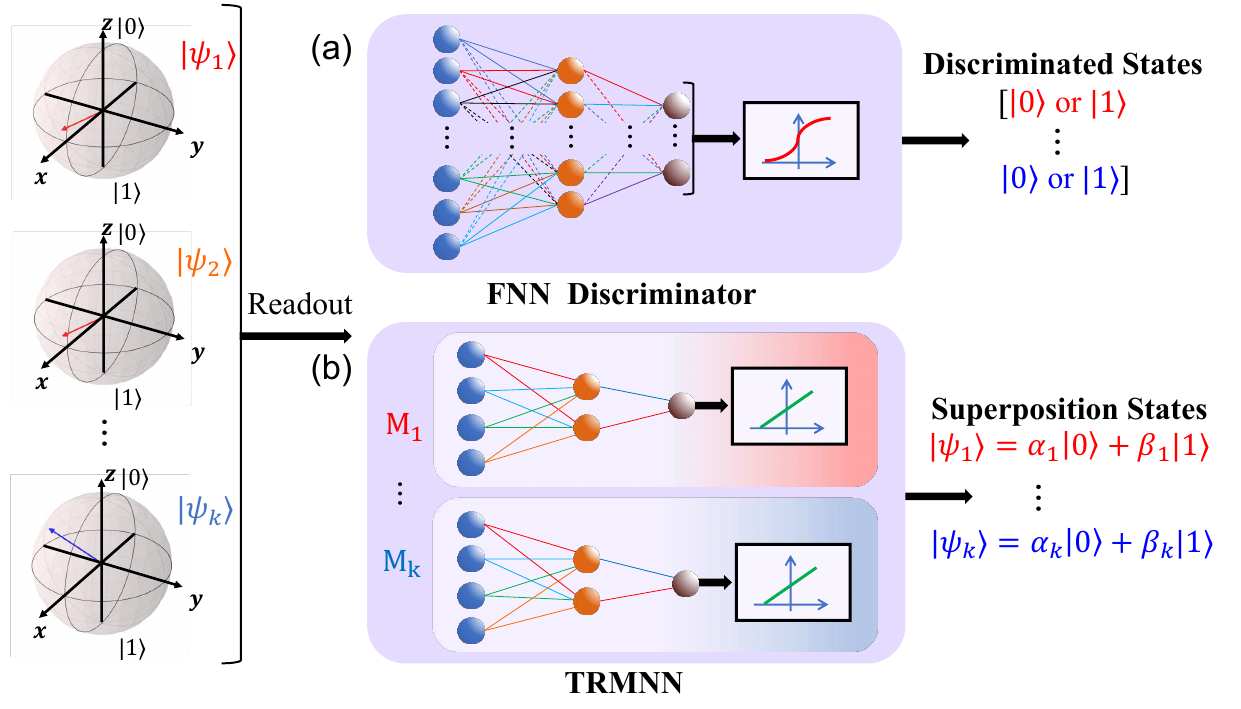}

\caption{(a) The neural network architecture of the feedforward neural network
(FNN) discriminator comprises a complex neural network designed for
discriminating between ground and excited states in multi-qubit systems.
The blue, orange, and brown circles represent the input, hidden, and
output layers, respectively. The softmax function is applied in the
output layer, depicted by a red line as a nonlinear activation function.
The softmax function acts as a binary classifier to enhance discrimination
between the two states. (b) TRMNN consists of multiple modular neural
networks ($M_{k}$). Each modular network is paired with a single
qubit. A probability estimation function in the output layer is represented
by a green line, indicating a linear activation function, and allowing
for the measurement of arbitrary superpositions.\protect\label{readout-arch}}
\end{figure}

The TRMNN is built upon a feedforward neural network (FNN) and undergoes
supervised learning using demodulated waveforms of
dispersive readout when the qubit is in superposition of the ground
and the excited states, see Fig.~\ref{readout-arch}.
The network is modulated in the sense that
each qubit in the system is paired with an identical network module
which is trained from data tagged either as the ground or the excited
waveform samples. Compared to raw readout demodulated in the phase space, the post-processing
neural network further enhances the readout fidelity through learned
recognition of the pure signals from the noises in the measurements.
The neural network essentially acquires node weights from reference
waveforms associated with ground and excited qubit states through
tagged data. A similarity score is subsequently generated through
a linear output activation function by comparing each sampled waveform
of unknown states to the reference waveforms.


\begin{figure}
\includegraphics[bb=14bp 0bp 585bp 425bp,clip,width=8.6cm]{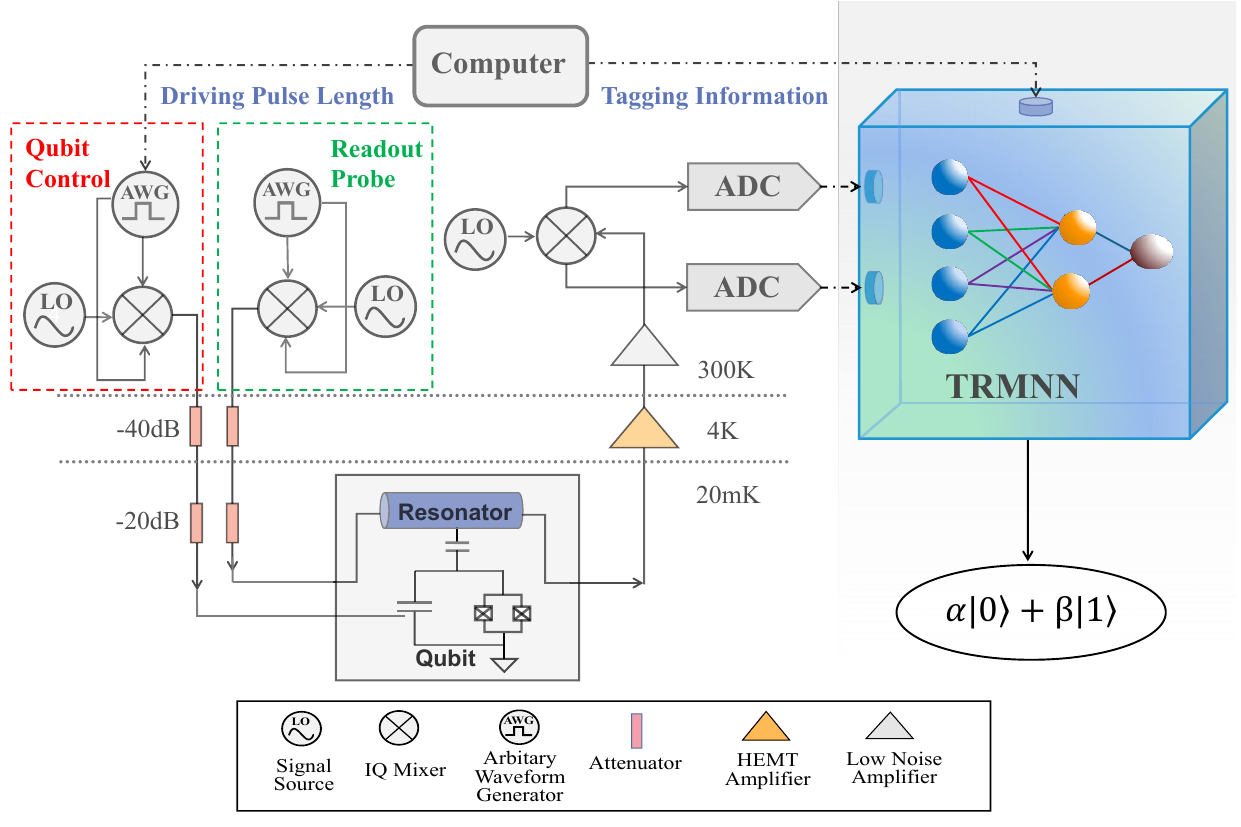}\caption{The system diagram illustrates the readout setup for a time-resolved
modulated neural network (TRMNN). Qubit control and readout probe
pulses are generated by arbitrary waveform generators (AWGs), upconverted
by IQ mixers, and directed to the testing qubit through attenuators.
The transmitted readout signal undergoes amplification, first by a
high electron-mobility transistor (HEMT) amplifier at a 4K stage and
then by a low-noise amplifier at room temperature. The TRMNN is implemented
after the analog-to-digital converters (ADCs) to analyze the IQ-demodulated
signal. A computer is utilized to send control commands to the AWG
and provide tagging information for the corresponding waveform to
the TRMNN during both the training and testing procedures. \protect\label{readout}}
\end{figure}

\section{Experiments and Results\protect\label{sec:Experiments-and-Results}}

We benchmark our neural network architecture with a single Xmon qubit coupled to a readout
resonator in a lattice of six superconducting qubits~\cite{Barends},
where the parameters of the testing qubit system are provided in Tab.~\ref{parameters}. The state information of the Xmon qubit is measured by probing a coupling resonator in the dispersive limit. The Hamiltonian of the qubit-resonator system is simplified as~\cite{Blais}
\begin{equation}
H_{\textrm{disp}}/\hbar=\left(\omega_{r}+\chi\sigma_{z}\right)\left(a^{\dagger}a+\frac{1}{2}\right)+\left(\omega_{q}+\chi\right)\frac{\sigma_{z}}{2},
\end{equation}
where $\omega_{r}$ denotes the bare frequency of the resonator, $\omega_{q}$
the qubit frequency, and $\chi=g^{2}/\Delta$ the state-dependent
frequency shift due to the coupling strength $g$ and the detuning
$\Delta=\omega_{r}-\omega_{q}$ between the qubit and the resonator.
Here, a change in the qubit state leads to an associated shift in
the resonator frequency through the term $\chi\sigma_{z}$, which
can be observed and quantified via a microwave probe signal~\citep{Blais}. 
In the experimental setup illustrated in Fig.~\ref{readout}, the
probing signal is shaped by an arbitrary waveform generator (AWG)
before reaching the target quantum chip. The response of the probing signal
is successively amplified by a high electron-mobility transistor (HEMT)
amplifier in low-temperature and a room-temperature low-noise amplifier.
The qubit state is encoded in both the phase and the amplitude of
the output probe signal. Considering a random noise term $N(t)$, the readout
signal can be written as

\begin{equation}
S(t)=S_{0}\cos\left(\omega_{\textrm{RO}}t+\theta_{\textrm{RO}}\right)+N(t),
\end{equation}
where $S_{0}$ and $\theta_{\textrm{RO}}$ are the state-dependent
amplitude and phase, respectively. The amplitude and phase are detected
through a heterodyne measurement using an analog In-phase and Quadrature
(IQ) mixer, which combines the readout signal $S(t)$ with a reference
local oscillator (LO) along two branches. 

\begin{figure}
\includegraphics[bb=0bp 20bp 595bp 465bp,clip,width=8.6cm]{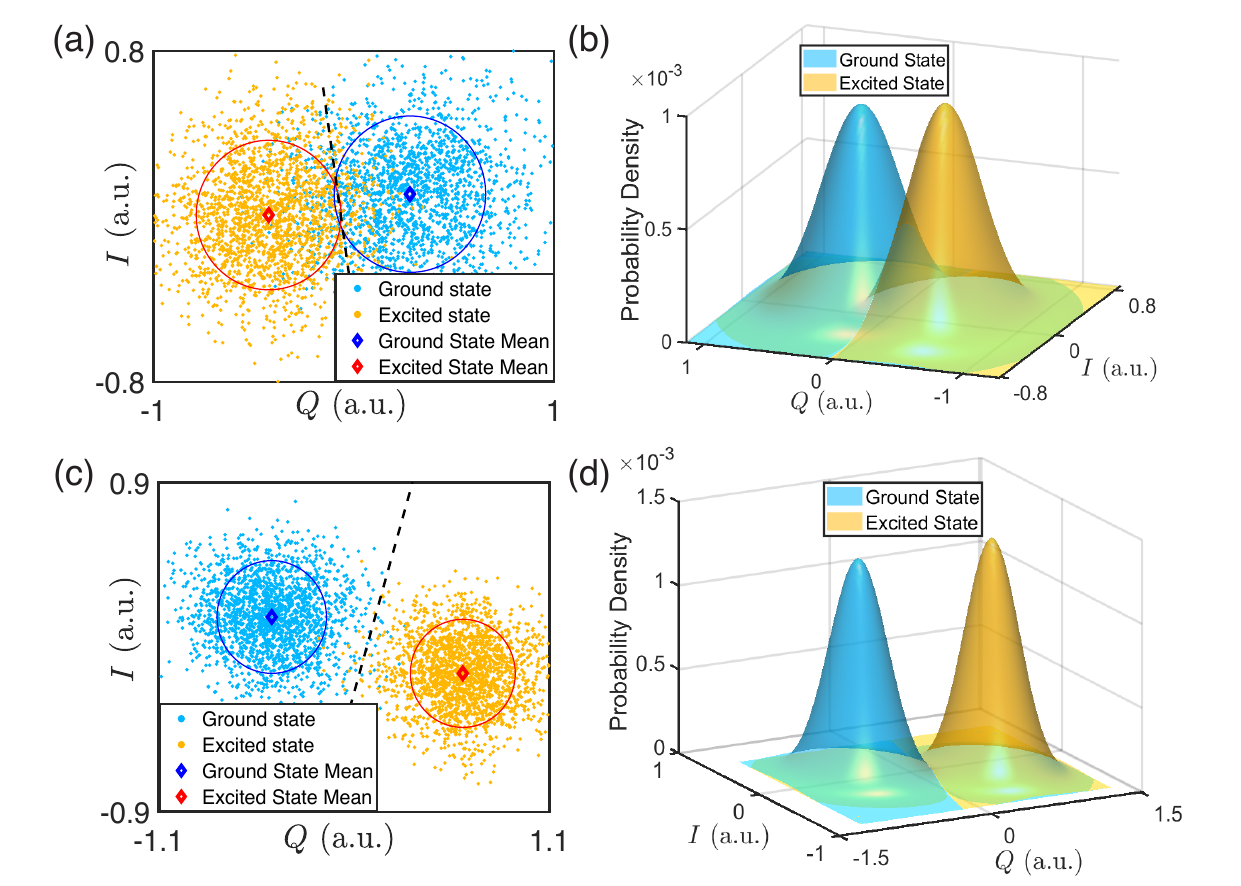}

\caption{(a) The raw readout data are plotted in the phase space when the qubit
is prepared in the ground and excited states in a low signal-to-noise
ratio (SNR) case, with 2000 shots per qubit state (fidelity $F=80.1$\%).
The circles in red and blue represent the variances for each state.
(b) Probability density regarded as Gaussian distributions with means
and variances extracted from (a) for both ground and excited states.
(c) The readout of the two states is plotted on the IQ plane in a
high SNR case, with 2000 shots per qubit state (fidelity $F=97.3$\%).
(d) Same as (b) but with means and variances extracted from (c).~\protect\label{IQ_plot}}
\end{figure}

Subsequently, the demodulated signals are sampled into a discrete-time
sequence by analog-to-digital converters (ADCs) in the I and Q branches
along with the noise. Each IQ pair measured is mapped to a specific
point in the phase space as shown in Fig.~\ref{IQ_plot}, where two
distinct clouds appear after a sufficient number of measurements.
While the separation between the clouds reflects signal strength,
the extension of each cloud shows the noise level $N(t)$. Subplots
(a) and (c) illustrate two typical detection scenarios: one with low
probe power ($-134$ dBm) and one with high probe power ($-131$ dBm). Along
with the increase in probe power, the signal-to-noise ratio (SNR)
is increased and thus the readout fidelity is enhanced from 80.1\%
to 97.3\%.

The noise effect embodied by the variance in the sampled data is persistent
in both low and high SNR scenarios. To mitigate this, neural network
methods have been widely adopted for recognizing categorized patterns and distilling
features from data containing noise~\cite{Badri,Xiao-Ping}. We
design the Time-Resolved Modulated Neural Network (TRMNN) to further
process the sampled raw data such that the readout fidelity can be
further increased, given the presence of noise.


Experimental assessments of qubit readout fidelity are conducted under
both low and high SNR conditions in different readout settings. The
readout performances of the raw readout, the FNN discriminator, and
TRMNN are presented to provide a comprehensive evaluation. In the
experiment, the dataset for each SNR setting comprises two subsets
of 5000 single-shot waveforms for the ground and excited states, respectively.
Within these subsets, 4000 sequences are allocated for training both
the FNN discriminator and TRMNN, while the remaining 1000 sequences
are reserved for testing.

The first evaluation of readout performance is the assignment fidelity~\cite{Heinsoo},
given by $F=1-[P(g|e)+P(e|g)]/2.$ In this equation, $P(g|e)$ ($P(e|g)$)
represents the probability of reading out the ground (excited) state
when the qubit is prepared in the excited (ground) state. The assignment
fidelity results for the three approaches are presented in Tab.~\ref{Assignment_Fidelity}.
In both high and low SNR cases, the fidelities of the FNN discriminator
and the TRMNN exhibit improvements compared to the raw readout. For
example, in the high SNR scenario, the two neural network-based approaches
show a similar improvement in fidelity, increasing from 97.3\% in
the raw readout case to 99.9\%. In the low SNR scenario, the fidelities
of the FNN discriminator and the TRMNN experience a notable improvement from 80.1\% to 98.5\%, more than an order of magnitude improvement in terms of infidelity reduction. In the measurement of
assignment fidelity, TRMNN consistently achieves similar performance
to the FNN discriminator, even without a binary classifier like the
softmax function in the neural network architecture.

\begin{table}
\begin{tabular}{|c|c|c|}
\hline 
$F$ & High SNR & Low SNR\tabularnewline
\hline 
\hline 
Raw Readout  & 97.3\% & 80.1\%\tabularnewline
\hline 
FNN Discriminator & 99.9\% & 98.6\%\tabularnewline
\hline 
TRMNN & 99.9\% & 98.5\%\tabularnewline
\hline 
\end{tabular}

\caption{The assignment fidelities for the raw readout, the FNN discriminator,
and TRMNN in the high and low SNR cases.\protect\label{Assignment_Fidelity}}
\end{table}

\begin{figure}
\includegraphics[width=8.6cm]{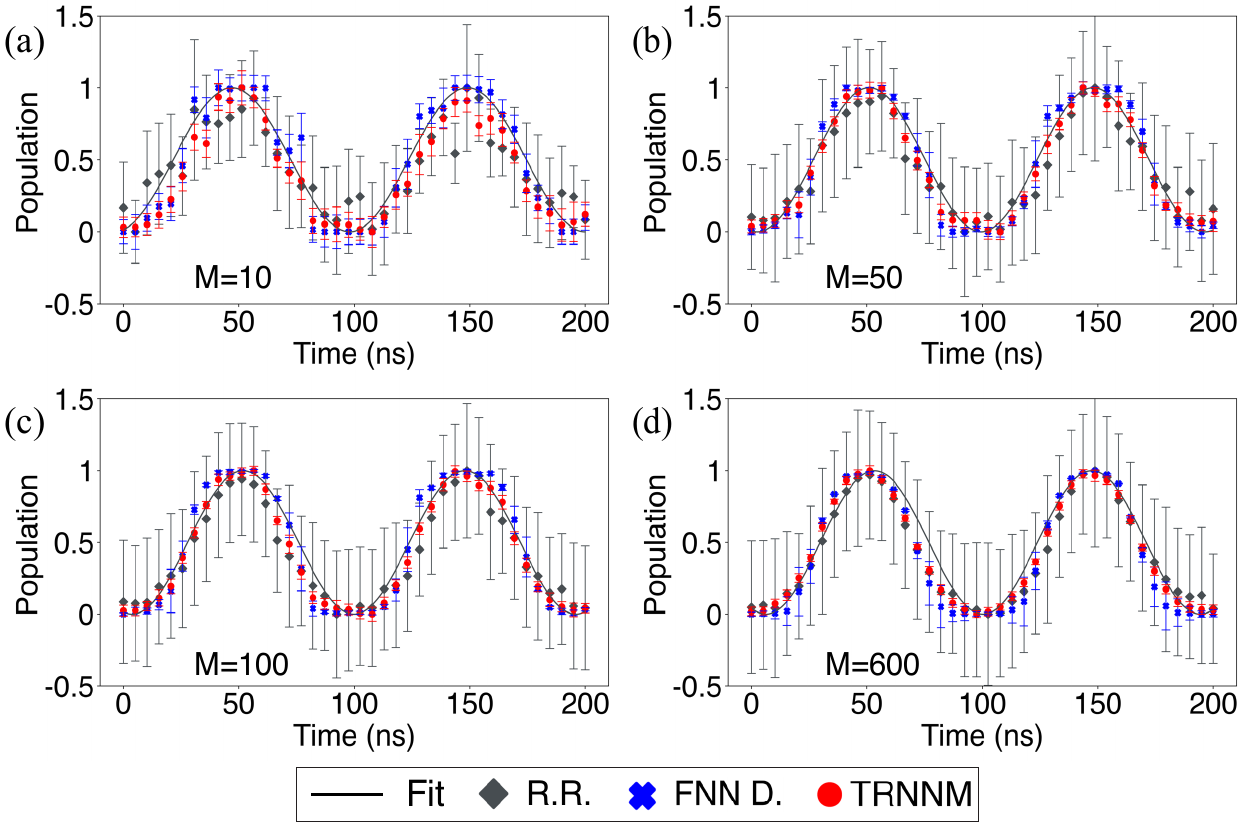}

\caption{Measurements of the excited state population for low SNR (fidelity
$F=80.1$\% at raw readout) over the duration of two Rabi oscillations
using raw readout (gray diamonds), FNN discriminator (blue crosses),
and TRMNN (red circles). Locations of the symbols indicate the statistical
average of (a) $M=10$, (b) $M=50$, (c) $M=100$, and (d) $M=600$
measurement counts at each time step over 200 ns, where the error
bar shows the Gaussian variance of the measurements. Under $M=600$,
the variances averaged over the entire time range are 1, 0.51, and
0.52 (a.u.) for the raw readout, the FNN discriminator, and TRMNN,
respectively.\protect\label{low_SNR}}
\end{figure}

\begin{figure}
\includegraphics[width=8.6cm]{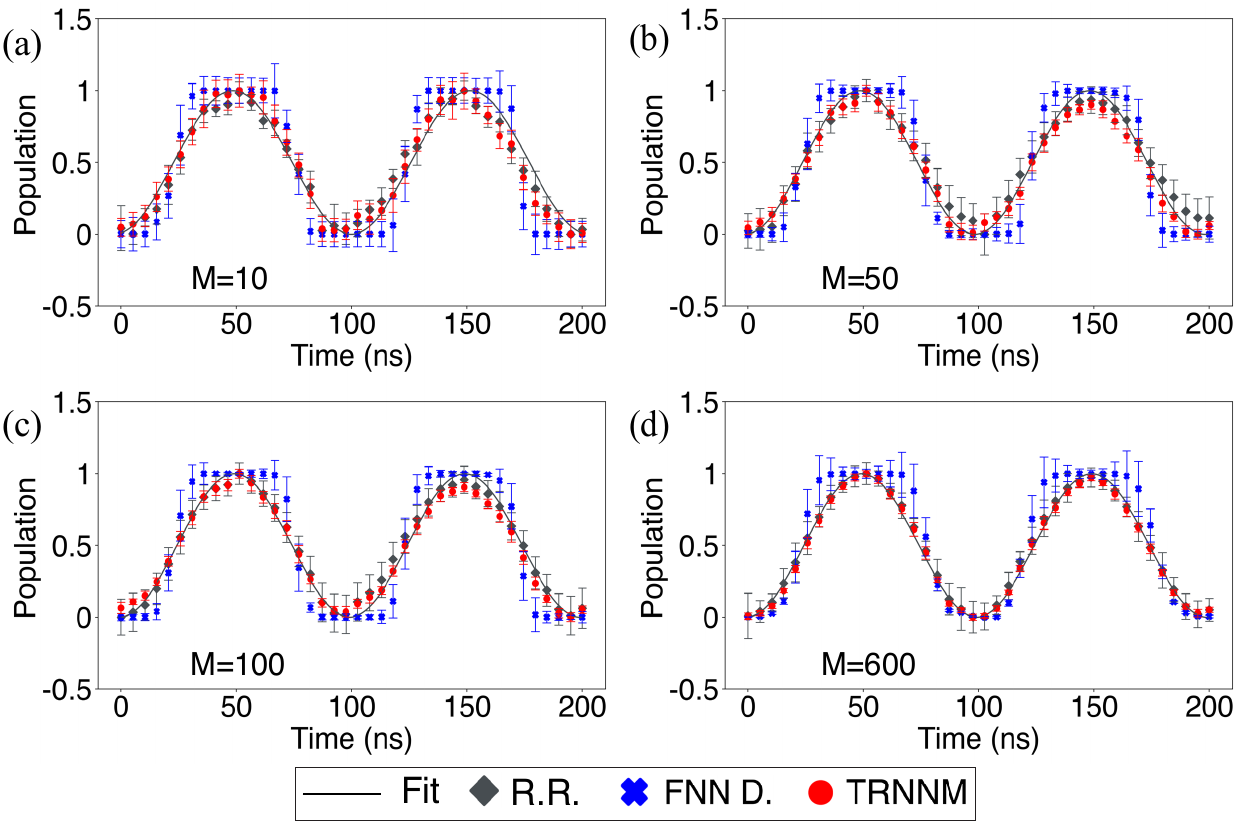}

\caption{Identical measurements as in Fig. 4 except for high SNR (fidelity
$F$=97.3\% at raw readout). The temporally averaged variances are
reduced to 0.15, 0.12, and 0.078 (a.u.) for the raw readout, the FNN
discriminator, and TRMNN, respectively. \protect\label{high_SNR}}
\end{figure}

The second evaluation is based on Rabi oscillation measurements to
compare the readout performance when the qubit is in an arbitrary
superposition state. Each Rabi oscillation measurement in the testing
dataset comprises 40 samples taken at a fixed time step within a 200~ns
interval, and the testing dataset includes 600 Rabi oscillation measurements.
Similar to the previous experiment, two distinct datasets for both
low and high SNR cases are recorded and analyzed by the trained neural
networks to observe their performances in different scenarios. The
measured oscillations of the excited state population with the Gaussian
variance are presented in Figs.~\ref{low_SNR} and ~\ref{high_SNR},
where the curves represent averages of sequences with 10, 50, 100,
and 600 Rabi oscillation measurements (M represents the number of
measurements), respectively. 

In Figs.~\ref{low_SNR} and ~\ref{high_SNR}, the results of the FNN
Discriminator exhibit oscillations in trapezoid shapes, distinguishing
them from the raw readout and TRMNN cases. This phenomenon arises from errors
introduced by the softmax function in the output layer of the network.
The softmax function tends to misclassify superposition states around
the ground and excited states as purely ground or excited states,
leading to enlargement around the peak and bottom of the curve. This
misclassification is also evident in the increased variance in the
FNN Discriminator curve around the region of the ground and excited
states.

The raw readout and TRMNN display similar results of Rabi oscillations
in both low and high SNR scenarios, making it challenging for a direct
evaluation based on the shapes of curves. However, concerning the
state variances, TRMNN shows a 48\% reduction from the raw readout
(1 to 0.52 and 0.15 to 0.078 a.u.) in the low and high SNR scenarios.
The reduction in variance suggests that neural network-based approaches
effectively reduce the impact of noise, aligning with the observed
enhancement in assignment fidelity.

Another evaluation for the Rabi oscillation measurement involves fitting
a sine function to the output curve and assessing the fitness since
ideal Rabi oscillations have a sine wave pattern with a decaying envelope
due to decoherence effects. Therefore, the fitness of a sine function
within a short time period can evaluate the accuracy of the readout
of Rabi oscillations, where the decoherence effect is not obvious.
Here a Rabi fidelity is defined as $F_{R}$, derived from the calculation
of the coefficient of determination~\cite{Barrett} applied to the
sine function fit as

\begin{equation}
F_{R}=1-\sum_{i=1}^{n}\frac{(y_{i}-f_{i})^{2}}{(y_{i}-\bar{y})^{2}},
\end{equation}
where $y_{i}$ represents the measured value, $f_{i}$ is the value
from the fit at each sampling point, respectively, and $\bar{y}$
is the average value of the measured curve. When the measured curve
perfectly aligns with the sine function fit, the value of $F_{R}$
equals unity.

\begin{table}
\begin{tabular}{|c|c|c|c|c|}
\hline 
$F_{R}$ & M=10 & M=50 & M=100 & M=600\tabularnewline
\hline 
\hline 
Raw Readout (Low SNR) & 0.849 & 0.928 & 0.947 & 0.955\tabularnewline
\hline 
FNN Discriminator (Low SNR) & 0.951 & 0.966 & 0.968 & 0.960\tabularnewline
\hline 
TRMNN (Low SNR) & 0.955 & 0.979 & 0.980 & 0.982\tabularnewline
\hline 
Raw Readout (High SNR) & 0.959 & 0.986 & 0.989 & 0.996\tabularnewline
\hline 
FNN Discriminator (High SNR) & 0.921 & 0.936 & 0.933 & 0.943\tabularnewline
\hline 
TRMNN (High SNR) & 0.984 & 0.985 & 0.986 & 0.993\tabularnewline
\hline 
\end{tabular}\caption{The Rabi fidelities by fitting the readout waveform with a sine function
for the raw readout, the FNN discriminator, and the TRMNN, in a high
and low SNR cases by averaging sequences of 10, 50, 100, and 600 times
of Rabi oscillations measurement. \protect\label{Rabi_Fidelity}}
\end{table}

Tab.~\ref{Rabi_Fidelity} provides a summary of the $F_{R}$ values
of the three approaches in the two SNR cases with different numbers
of averaged shots. In the low SNR case, TRMNN consistently exhibits
the highest Rabi fidelity, irrespective of the number of averaged
shots. This outcome reinforces the earlier observation that TRMNN
effectively mitigates the adverse effects of noise and enhances overall
readout fidelity compared to the raw readout. Meanwhile, the FNN Discriminator
has the second-highest $F_{R}$ value, which is consistent with the
trapezoid shape error introduced by the softmax function. In the high
SNR case, the results indicate that raw readout and TRMNN consistently
obtain well-fitting sine outputs in high SNR scenarios. Nevertheless,
TRMNN attains the highest result with a small number of measurements
($M=10$), highlighting its potential application in single-shot readouts
for state tomography.

\section{Conclusion\protect\label{sec:Conclusion}}

In this study, we introduce a time-resolved modulated neural network
specifically designed for high-fidelity qubit state tomography. This
approach effectively promotes readout fidelity by reducing noise variances
in readout waveforms. Its modulated nature encourages adaptable scalability
for the circuit topology under detection and its linear-correlation
nature permits time-domain measurements at desired time step and duration.
The proposed TRMNN showcases its advantages not only in
effectively classifying states into both the ground and excited states
but also in achieving a high level of accuracy in reading arbitrary
superposition states.
Our experimental results have verified the performance gain, proving
the usefulness of this novel architecture.
Moreover, the proposed modular network architecture
ensures straightforward scalability to handle potential expansions
in the size of qubits.
It expands the scope of machine learning within the field of research of superconducting qubits
and quantum computation in general. 

\begin{acknowledgments}
H. I. thanks the support
of FDCT of Macau under grants 0015/2021/AGJ and 006/2022/ALC.
Y. Z. thanks the support of NSFC of China under grant 12174178
and the Science, Technology and Innovation Commission of Shenzhen
Municipality under grant KQTD20210811090049034.
\end{acknowledgments}

\section{Appendix}

\subsection{Network architecture }

Neural networks can be classified into various architectural categories,
including feedforward neural networks (FNN)~\cite{Svozil} and recurrent
neural networks (RNN)~\cite{Schuster}, based on the structure of
layers and the connections between neurons. Previous research has
demonstrated that the fully connected FNN architecture improves readout
fidelity as a state discriminator for multiple qubits~\cite{Lienhard}.
The architecture of an FNN discriminator is illustrated in Fig.~\ref{readout-arch}(a),
where a single neural network is implemented to discriminate between
the ground and excited states for all measured qubits. Hence, along
with the number of qubits, the network complexity (both the number
of hidden layers and the necessary size of training sets) scales up,
challenging both readout accuracy and scalability simultaneously.

In contrast, the TRMNN is a network cluster comprising as many FNNs
and output function instances as the number of qubits: each qubit
matches with a network module with identical internal architecture,
as illustrated in Fig.~\ref{readout-arch}(b). Each module includes
an input layer, an output layer, and three hidden layers consisting
of 900, 250, and 50 neurons, respectively, in between. Intra-layer
connections are furnished by rectified linear unit (ReLU)~\cite{key-2}
for activation, whereas the output layer is connected to a linear
output activation function often used in regression prediction~\cite{Imrie}
for probability estimation. State tomography for qubits here is akin
to linear regression analysis: the neural network takes a sampled
waveform as the input to generate in an output neuron a score over
the interval $\left[0,1\right]$ based on its similarity or distance
to the preset ground or excited state waveforms defined as 0 and 1,
respectively.

FNN equipped with a discriminator incorporates a softmax function
in the output layer, which is a nonlinear output activation function
widely used for classification problems~\cite{key-1}. It acts as
a binary classifier in enhancing distinguishability between ground
and excited state waveforms but has the drawback of misclassification
when the qubit superposition is close to but not identical to either
the ground or the excited state. TRMNN, on the other hand, generates
a direct output linearly correlated with the population inversion
without additional enhancement, thereby enhancing the readout fidelity
for all superposition states. The trained network focuses on mitigating
the noise effect rather than improving binary distinguishability.
Moreover, the fixed architecture from the input layer to the output
function of each module permits straightforward scalability by expanding
the module number when the number of qubits is increased. Through
parallel computational expansion, the increase in the number of network
modules has a modest impact on computational resources.

\subsection{Dataset preparation and network training }

TRMNN requires training through supervised learning before it can
effectively operate. The training dataset comprises fixed-length sampled
waveforms corresponding to the ground and the excited states of its
paired qubit. The experimental results shown below are obtained from
4,000 probe shots each for either the ground or the excited state,
where each shot consists of a sequence of 2,000 sampled IQ point pairs
in one microsecond.

During network training, the qubit is initially prepared according
to the target state expected. To be exact, as illustrated in Fig.~\ref{readout},
the computer issues commands to the AWG such that it either sends
a $\pi$-pulse to invert the qubit or remains silent to let the qubit
stay at the ground state. Simultaneously, tagging information of the
corresponding state is transmitted to the TRMNN to associate each
emitted waveform sampled on the ADC with each tag.

The connection weights between the neurons are adjusted by minimizing
the error in the output compared to the expected result. To optimize
training performance, a validation-training set ratio of 0.2 is employed,
utilizing the Adam optimizer~\cite{Kingma} with mean squared error
as the loss function. Then, the trained TRMNN inherently distills
waveform differences in the tagged training set and averages out the
random noise in the connection weights. Consequently, the output neuron
accumulates a weight proportional to the similarity score where the
noise has been filtered out during the accumulation. When this weight
is read through the output linear activation function, its variance
and fidelity would be improved over the pre-network raw readout.

After the training is completed, TRMNN receives the testing waveform
of an arbitrary superposition state, which can consist either of a
single shot or a set of repetitive measurements. By comparing it to
the trained waveforms, the neural network generates a similarity score
and translates it into a qubit population. In the case of repetitive
measurements, TRMNN directly produces an averaged result and variance
for the testing set, removing the averaging step necessary for raw
IQ quadratures readout.

The time resolution of TRMNN is accomplished by associating the testing
waveform with the qubit control pulse length. Particularly, the readout
results can be plotted against arbitrary driving pulse lengths, enabling
comprehensive time-domain state tomography over desired durations.
Therefore, it can capture Rabi oscillations at specific time steps,
bypassing the limitation of discriminator-type networks.

\begin{table}
\begin{tabular}{|c|c|}
\hline 
\multicolumn{2}{|c|}{System parameters}\tabularnewline
\hline 
\hline 
Xmon qubit frequency & 3.842 GHz\tabularnewline
\hline 
Readout resonator frequency & 5.331 GHz\tabularnewline
\hline 
Coupling strength & 85 MHz\tabularnewline
\hline 
Readout resonator linewidth & 1.1 MHz\tabularnewline
\hline 
$T_{1}$ & 26 us\tabularnewline
\hline 
$T_{2}$ & 5 us\tabularnewline
\hline 
\end{tabular}

\caption{The parameters in the testing system of a single xmon qubit coupled
to a readout resonator. \protect\label{parameters}}
\end{table}


\begin{thebibliography}{99}

\bibitem{Montavon}G. Montavon, M. Rupp, V. Gobre, A. Vazquez-Mayagoitia,
K. Hansen, A. Tkatchenko, K.-R. Müller, and O. Anatole von Lilienfeld,
New J. Phys. \textbf{15}, 095003 (2013).

\bibitem{Tiunov}E. S. Tiunov, V. V. Tiunova, A. E. Ulanov, A. I.
Lvovsky, and A. K. Fedorov, Optica \textbf{7}, 448 (2020).

\bibitem{Ahmed}S. Ahmed, C. S. Mu\"noz, F. Nori, and A. F. Kockum,
Phys. Rev. Lett. \textbf{127}, 140502 (2021).

\bibitem{Seif}A. Seif, K. A. Landsman, N. M. Linke, C. Figgatt, C.
Monroe, and M. Hafezi, J. Phys. B: At., Mol. Opt. Phys. \textbf{51},
174006 (2018).

\bibitem{Ding}Z.-H. Ding, J.-M. Cui, Y.-F. Huang, C.-F. Li, T. Tu,
and G.-C. Guo, Phys. Rev. Appl. \textbf{12}, 014038 (2019).

\bibitem{Convy}I. Convy, H. Liao, S. Zhang, S. Patel, W. P. Livingston,
H. N. Nguyen, I. Siddiqi, and K. B. Whaley, New J. Phys. \textbf{24},
063019 (2022).

\bibitem{Koolstra}G. Koolstra et al., Phys. Rev. X \textbf{12}, 031017
(2022).

\bibitem{Cao}S. Cao et al., arXiv preprint arXiv:2402.09532 
(2024).

\bibitem{Arute}F. Arute et al., Nature \textbf{574}, 505 (2019).

\bibitem{Clarke}J. Clarke and F. K. Wilhelm, Nature \textbf{453},
1031 (2008).

\bibitem{Ladd}T. D. Ladd, F. Jelezko, R. Laflamme, Y. Nakamura, C.
Monroe, and J. L. O'Brien, Nature \textbf{464}, 45 (2010).

\bibitem{google2023suppressing}Google Quantum AI, Nature, \textbf{614}, 676--681 (2023).

\bibitem{Borjans}F. Borjans, X. Mi, and J. R. Petta, Phys. Rev. Appl.
\textbf{15}, 044052 (2021).

\bibitem{Myerson}A. H. Myerson et al., Phys. Rev. Lett. \textbf{100},
200502 (2008).

\bibitem{Robledo}L. Robledo, L. Childress, H. Bernien, B. Hensen,
P. F. Alkemade, and R. Hanson, Nature \textbf{477}, 574 (2011).

\bibitem{Gambetta}J. M. Gambetta, J. M. Chow, and M. Steffen, Npj
Quantum Inf. \textbf{3}, 2 (2017).

\bibitem{Lupa=00015Fcu}A. Lupa\c{s}cu, S. Saito, T. Picot, P. C.
de Groot, C. J. P. M. Harmans, and J. E. Mooij, Nat. Phys. \textbf{3},
119 (2007).

\bibitem{Siddiqi}I. Siddiqi, R. Vijay, M. Metcalfe, E. Boaknin, L.
Frunzio, R. J. Schoelkopf, and M. H. Devoret, Phys. Rev. B \textbf{73},
054510 (2006).

\bibitem{Wang}Y. Wang, Z. You, and H. Ian, AVS Quantum Sci. \textbf{5}
(2023).

\bibitem{Klimov}P. V. Klimov et al., Phys. Rev. Lett. \textbf{121},
090502 (2018).

\bibitem{Martinis}J. M. Martinis, S. Nam, J. Aumentado, K. M. Lang,
and C. Urbina, Phys. Rev. B \textbf{67}, 094510 (2003).

\bibitem{Tripathi}V. Tripathi, H. Chen, M. Khezri, K.-W. Yip, E.
M. Levenson-Falk, and D. A. Lidar, Phys. Rev. Appl. \textbf{18}, 024068
(2022).

\bibitem{Mundada}P. Mundada, G. Zhang, T. Hazard, and A. Houck, Phys.
Rev. Appl. \textbf{12}, 054023 (2019).

\bibitem{Pitsun}D. Pitsun et al., Phys. Rev. Appl. \textbf{14}, 054059
(2020).

\bibitem{Walter}T. Walter et al., Phys. Rev. Appl. \textbf{7}, 054020
(2017).

\bibitem{Heisoo}J. Heisoo, et al., Phys. Rev. Appl. \textbf{10},
03404 (2018).

\bibitem{Boissonneault}M. Boissonneault, J. M. Gambetta, and A. Blais,
Phys. Rev. Lett. \textbf{105}, 100504 (2010).

\bibitem{Hornik}K. Hornik, Neural Netw. \textbf{4}, 251 (1991).

\bibitem{Cybenko}G. Cybenko, Math. Control. Signals, Syst. \textbf{2},
303 (1989).

\bibitem{Duan}P. Duan, Z.-F. Chen, Q. Zhou, W.-C. Kong, H.-F. Zhang,
and G.-P. Guo, Phys. Rev. Appl. \textbf{16}, 024063 (2021).


\bibitem{Lienhard}B. Lienhard et al., Phys. Rev. Appl. \textbf{17},
014024 (2022).

\bibitem{Magesan}E. Magesan, J. M. Gambetta, A. D. C\'orcoles, and
J. M. Chow, Phys. Rev. Lett. \textbf{114}, 200501 (2015).

\bibitem{Barends}R. Barends et al., Phys. Rev. Lett. \textbf{111},
080502 (2013).

\bibitem{Blais}A. Blais, R.-S. Huang, A. Wallraff, S. M. Girvin,
and R. J. Schoelkopf, Phys. Rev. A \textbf{69}, 062320 (2004).

\bibitem{Badri}L. Badri, Int. Arab J. Inf. Technol. \textbf{7}, 289
(2010).

\bibitem{Xiao-Ping}Z. Xiao-Ping, IEEE trans. neural netw. \textbf{12},
567 (2001).



\bibitem{Heinsoo}J. Heinsoo et al., Phys. Rev. Appl. \textbf{10},
034040 (2018).

\bibitem{Barrett}J. P. Barrett, Am. Stat. \textbf{28}, 19 (1974).


\bibitem{Svozil}D. Svozil, V. Kvasnicka, and J. Pospichal, Chemom.
Intell. Lab. Syst. \textbf{39}, 43 (1997).

\bibitem{Schuster}M. Schuster and K. K. Paliwal, IEEE Trans. Signal
Process. \textbf{45}, 2673 (1997).

\bibitem{key-2}J. Schmidt-Hieber, Ann. Stat. \textbf{48}, 1875 (2020).

\bibitem{Imrie}C. Imrie, S. Durucan, and A. Korre, J. Hydrol. \textbf{233},
138 (2000).

\bibitem{key-1}M. Jiang, Y. Liang, X. Feng, X. Fan, Z. Pei, Y. Xue,
and R. Guan, Neural. Comput. Appl. \textbf{29}, 61 (2018).

\bibitem{Kingma}D. P. Kingma and J. Ba, arXiv preprint arXiv:1412.6980
(2014).




\end{thebibliography}
\end{document}